\documentstyle[psfig,conf_iap,10pt]{article}
\begin{document}
\heading{Imaging the z=0.9 Absorbing Cloud toward 1830-211} 
\par\medskip\noindent
\author{
C.L. Carilli$^1$, K.M. Menten$^{2}$, M.J. Reid$^{3}$,
M.P. Rupen$^1$, and M. Claussen$^1$
}
\address{
National Radio Astronomy Observatory, Socorro, NM, 87801, U.S.A.
}
\address{
Max Planck Institut f\"ur Radio Astronomie, Bonn, Germany
}
\address{
Center for Astrophysics, Cambridge, MA, 02138
}

\begin{abstract}
We present spectroscopic  imaging observations
of the molecular and HI 21cm absorbing cloud at z = 0.885 toward 
the `Einstein ring' radio source PKS 1830-211. 
We derive a cloud size between 10 h$^{-1}$ pc and 600 h$^{-1}$ pc, and
M(H$_2$) $>$ 2.6$\times$10$^{4}$ h$^{-2}$ M$_\odot$.  
The temperature of the ambient radiation field
is 4.5$^{+1.5}_{-0.6}$ K, consistent with the
microwave background temperature at z = 0.885. The velocity difference
for absorption on opposite sides of the ring is -146 km
s$^{-1}$, which is consistent with the galaxian rotation velocity
derived from gravitational lens models. 
\end{abstract}

Wiklind and Combes \cite{WC1} have discovered a
strong molecular absorption line system toward the `Einstein ring'
radio source 1830-211 at z = 0.885 with 
N(H$_2$) = 3$\times$10$^{22}$ cm$^{-2}$ \cite{WC2} \cite{FRYE}.
We have begun an extensive program of imaging the pc-
and kpc-scale  structures in this absorbing cloud in many molecular 
species and transitions in order to study the astrochemistry, physical
conditions, structures, and dynamics of the dense ISM in this system.

The VLA 47 GHz image and spectra of the redshifted  
HCN(1-0) absorption toward 1830-211
are shown in Fig. 1. There is strong absorption seen toward the SW
radio component at z = 0.88582 with a peak optical depth of 2.5, 
and no absorption toward the NE component
at this redshift to an optical depth limit of 0.012 (3$\sigma$). 
There is a much weaker absorption line seen only toward the NE
radio component at  z = 0.88491 ~(= -146 km s$^{-1}$ relative to the
strong line), with a peak optical depth of 0.04.  The
fact that the   absorption along the line of sight to the SW
component is different from that toward the NE component
implies that this must be absorption by gas in  
a cosmologically intervening galaxy (presumably the lens), and not
at the radio source redshift. Both the NE and SW  radio components are 
spatially extended, and we can set a limit of 0.3 (3$\sigma$) to the
optical depth  toward the `tail' of the SW component, implying an
upper limit to the main cloud size of 600 h$^{-1}$ pc.

The VLBA image at 24 GHz and spectra  
of redshifted  HC$_3$N(5-4) absorption 
are shown in Fig. 2. The continuum source shows a possible core-jet
extending $\approx$ 2 mas to the northwest. The spectra show an
increase in absorbed flux density going to coarser resolution, 
suggesting a lower limit to the cloud size of 2.5 mas =
10 h$^{-1}$ pc, although observations of HC$_3$N(3-2) absorption
indicate  that there may be spatial sub-structure with velocity.
The upper limit to the
volume averaged H$_2$ density is $\sim 1000$ cm$^{-3}$, and the lower limit
to the mass is: M(H$_2$) $>$ 2.6$\times$10$^{4}$  h$^{-2}$
M$_\odot$. These values are comparable to those found for giant
molecular cloud complexes in our own galaxy \cite{VDIS}. 

From VLA observations of HC$_3$N (3-2) and (5-4)
we derive a temperature for the ambient radiation field
of 4.5$^{+1.5}_{-0.6}$ K, consistent with the temperature
of the microwave background at z = 0.885.
Absorption is seen by free radicals (C$_3$H$_2$), and by molecular
ions (HCO$^+$), characteristic of translucent Galactic molecular
clouds \cite{VDIS}.
The absorption line
velocity structure varies significantly between different molecular
species (Fig. 3b). We derive a Carbon isotope ratio in the molecular
gas phase of: ${C^{12}}\over{C^{13}}$ = 35, which is consistent
with significant fractionation, and a mature  gas phase chemistry
\cite{WC2}. We do not detect deuterated HCN,
leading to the limit: ${HCN}\over{DCN}$ $>$ 200, suggesting a
molecular formation temperature $\ge$ 20 K \cite{WOOT}.

Neutral hydrogen 21cm absorption has been detected toward 1830-211 \cite{MCM}
(Fig. 3). Absorption is detected  at both z =
0.88582 with N(HI) $\approx$ 0.5$\times$10$^{19}$ ${T_s}\over{f}$ cm$^{-2}$
and at z = 0.88491 with N(HI) 
$\approx$ 1.0$\times$10$^{19}$ ${T_s}\over{f}$ cm$^{-2}$, where $T_s$ is the
spin temperature and $f$ is the covering factor \cite{CHEN}. 
While the HI 21cm observations do not spatially resolve the radio
continuum source, we can infer from the molecular imaging that
the HI absorption at z = 0.88491 is toward the NE radio component,
while that  at z = 0.88582 is toward the SW component.
Assuming a Galactic dust-to-gas ratio  implies rest frame visual
extinctions $\ge$ 1 toward both radio components, for
$T_s$ $>$ 100 K.  However, the ratio of H$_2$ to HI column densities
differs by about two orders of magnitude between the two absorbing
systems (assuming similar $T_s$ values), hence one might also expect a
variable dust-to-gas ratio \cite{WC2}.

The velocity difference between the absorption toward the NE and SW
components can be used to check the gravitational  lens model by
assuming Keplerian rotation \cite{WC2}. From the lens model
of Nair etal. \cite{NAIR}, we calculate an expected velocity
difference between the two lines-of-sight of 144 km s$^{-1}$, under
the (very uncertain) assumption that the inclination 
angle of the lensing galaxy is given by the ellipticity of the lens mass
distribution. This is fortuitously close to
the observed velocity separation of 146 km s$^{-1}$. 
High resolution infrared imaging is required to constrain the
inclination angle of the lens, and thereby allow for a fundamental
check on gravitational lens theory.

\acknowledgements{The National Radio Astronomy Observatory is operated
by Associated Universities Inc., under contract with the National
Science Foundation }

\clearpage
\newpage

\begin{figure}
\psfig{figure=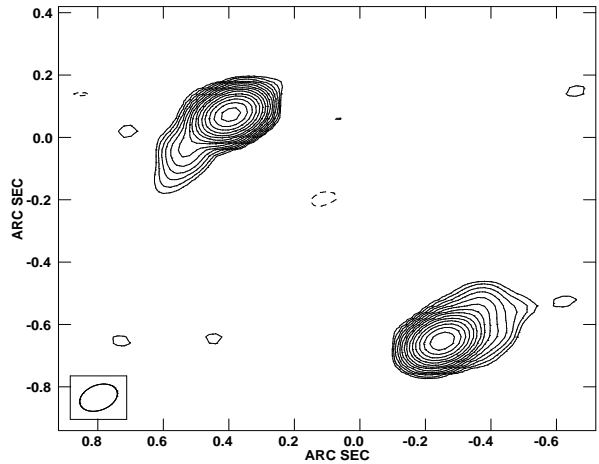,width=8cm,height=8cm}
\vspace{-1cm}
\psfig{figure=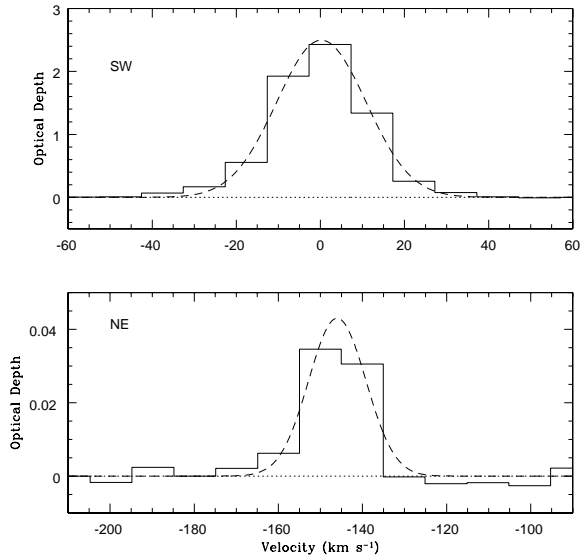,width=8cm,height=8cm}
\vspace{-0.5cm}
\caption{The upper frame shows the continuum image of 1830-211 
made with the VLA at 47 GHz with a resolution of 0.1$''$. 
The contour levels are a geometric progression
in square root two, and the first level is 7.5 mJy/beam. 
The lower figure shows redshifted HCN(1-0) absorption spectra toward
the two peaks in the continuum image (designated SW and NE).
Zero velocity corresponds to z$_\odot$ = 0.88582.
The spectra have been converted to optical depth using the
continuum surface brightness at each position.
The upper spectrum is at the peak surface brightness of the SW radio
component, where F$_\nu$ = 0.91 Jy/beam. The dash line 
is a Gaussian fit to the data, with a 
peak optical depth of 2.5, and FWHM = 25$\pm$5 km s$^{-1}$ centered at
0 km s$^{-1}$. The lower spectrum is toward the NE radio
component, where F$_\nu$ = 1.1 Jy/beam, and the Gaussian Fit
parameters are:  peak optical depth of 0.043 and  FWHM = 16$\pm$5 km s$^{-1}$
centered at -146$\pm$1 km s$^{-1}$. 
}
\end{figure}

\clearpage
\newpage

\begin{figure}
\vspace{-1.5cm}
\hspace{4.0cm}
\psfig{figure=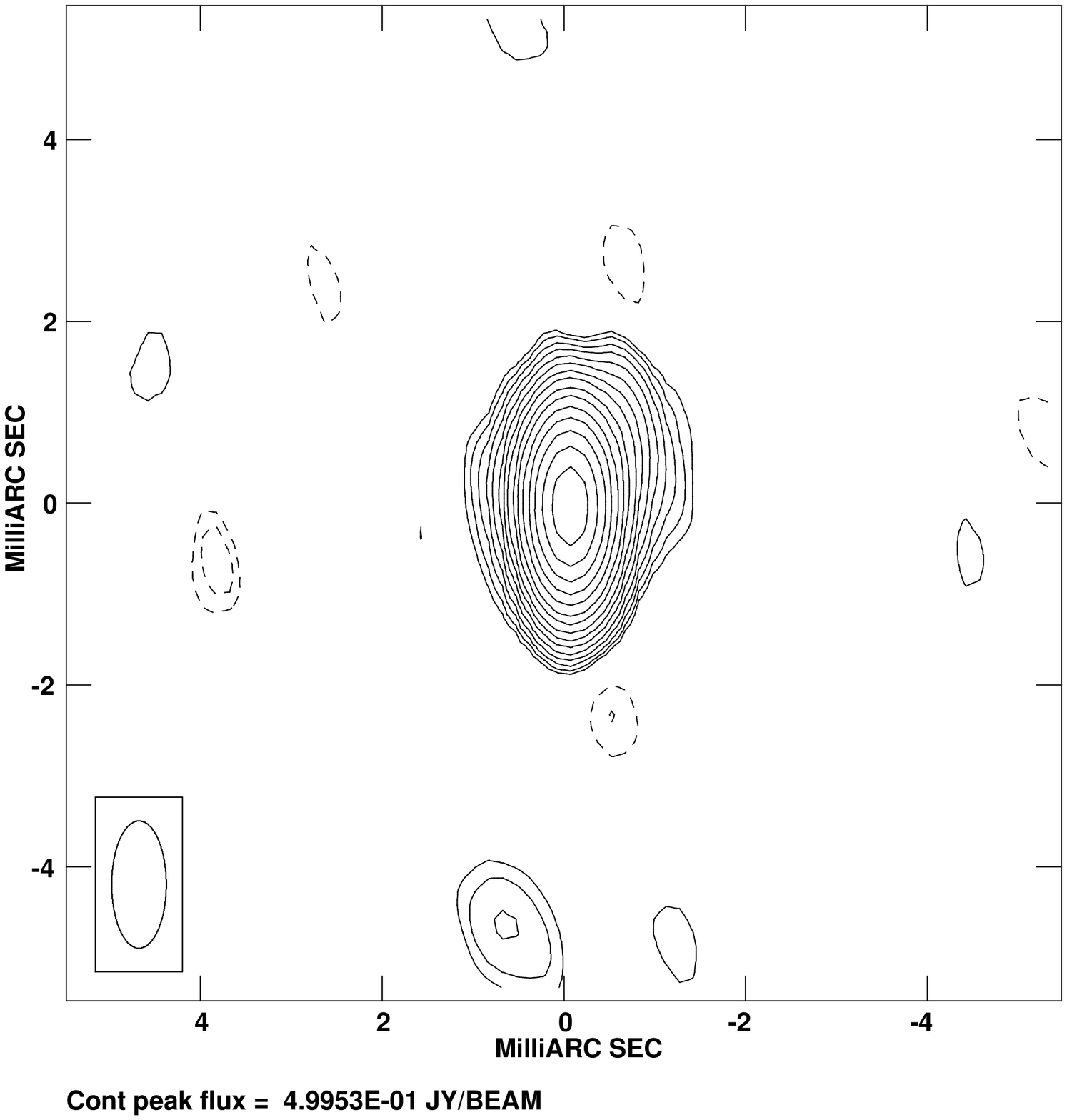,width=6cm,height=6cm}
\vspace{-6cm}
\hspace{7.5cm}
\psfig{figure=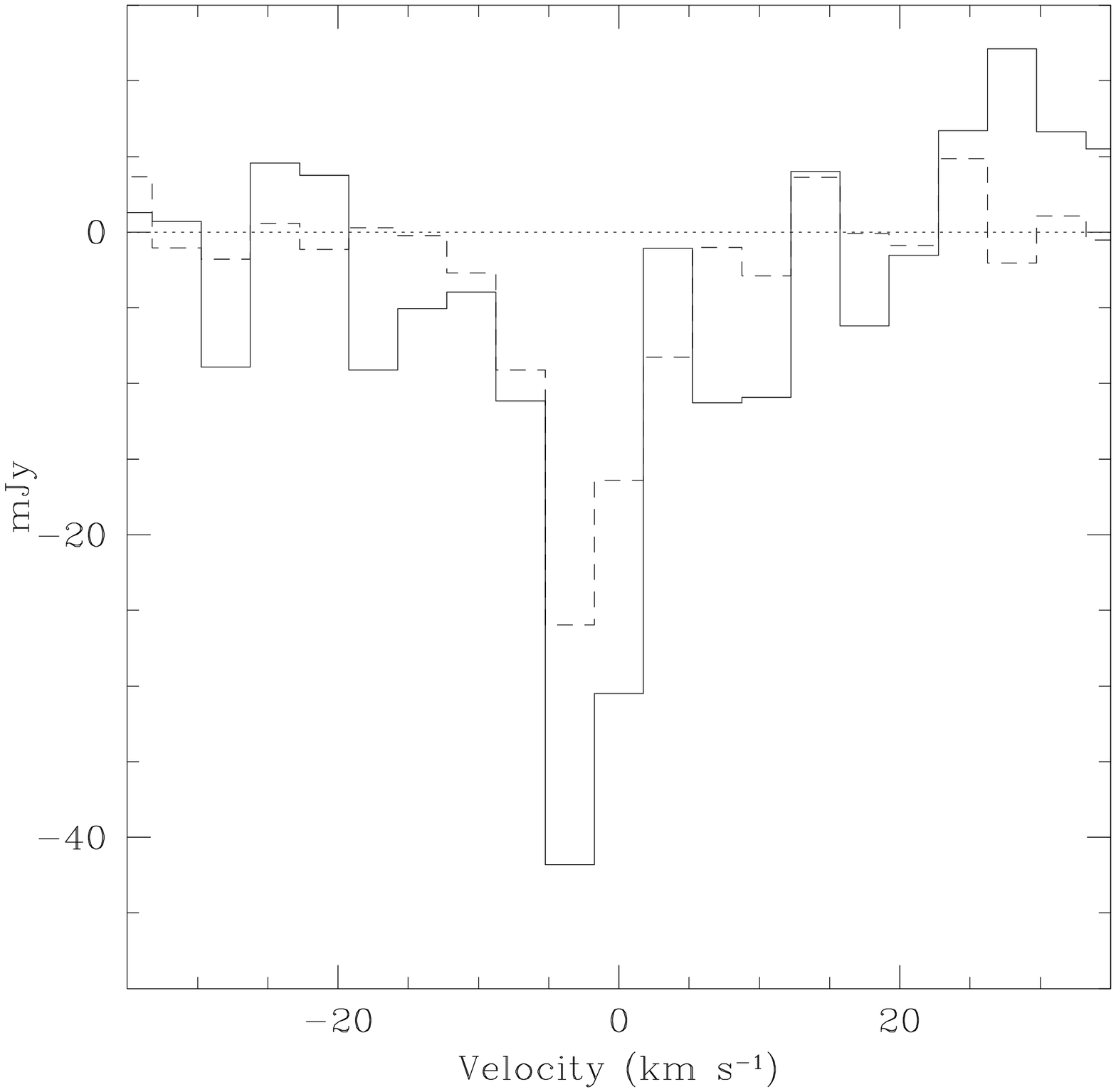,width=5cm,height=6cm}
\vspace{-0.4cm}
\caption{The figure on the right shows the continuum image of the
SW component of 1830-211 at 24 GHz made with the
VLBA  at 1 mas resolution. Contouring
is the same as Fig. 1, with the first level = 3 mJy/beam. 
The figure on the left shows  VLBA spectra of redshifted
HC$_3$N(5-4)  absorption toward the SW component. 
Zero velocity corresponds to
z$_\odot$ = 0.88582. The dash line is a spectrum at the peak surface
brightness  at 1mas resolution.  The solid line is the same spectrum,
but now at 2.5mas resolution. The peak line flux density increases from 
26$\pm$4 mJy at 1 mas resolution to 42$\pm$8 mJy at 2.5 mas resolution,
suggesting a lower limit to the cloud size of 2.5mas. 
}
\end{figure}

\begin{figure}
\vspace{-0.6cm}
\hspace{4.5cm}
\psfig{figure=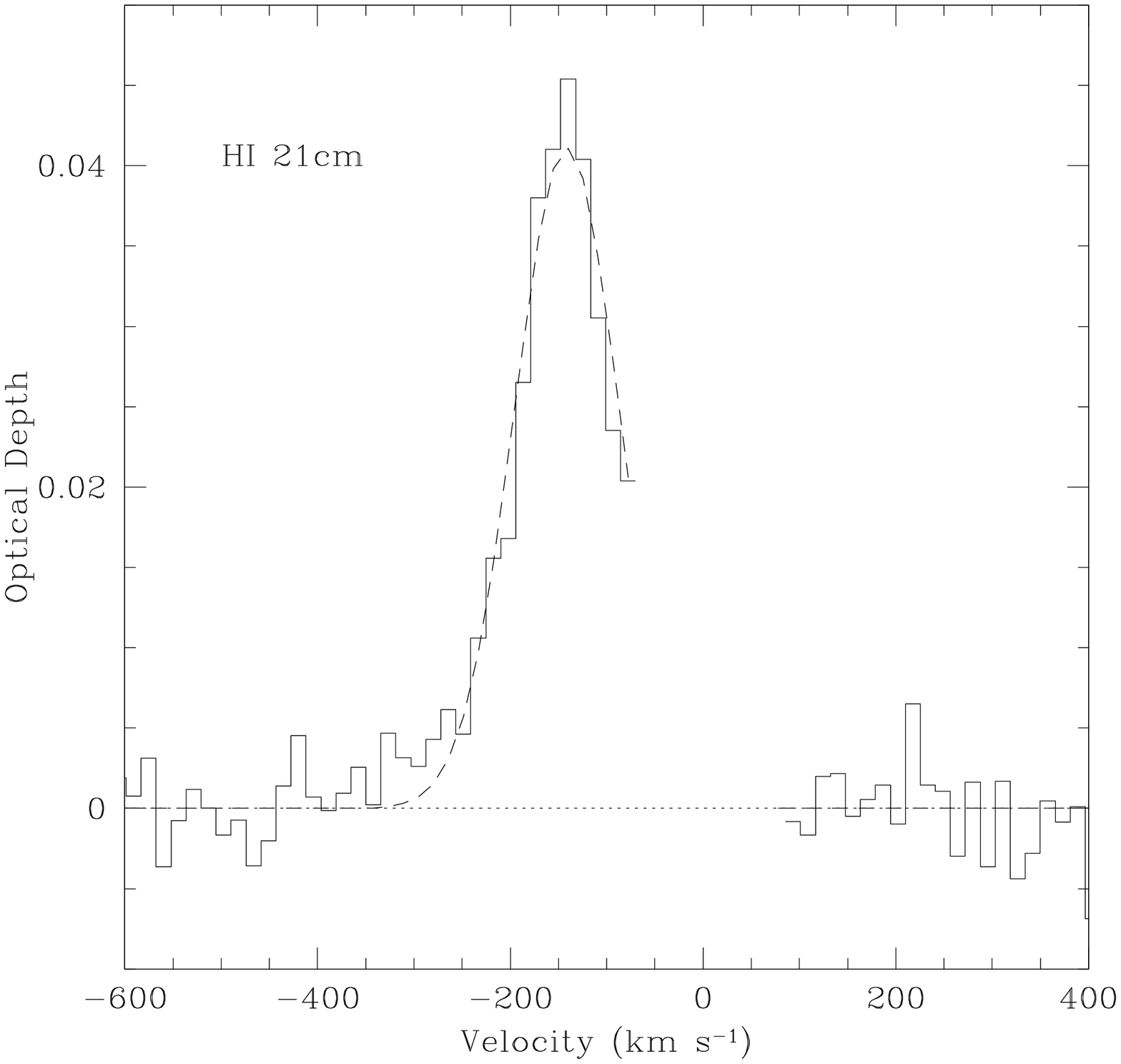,width=6cm,height=6cm}
\vspace{-6cm}
\hspace{7.5cm}
\psfig{figure=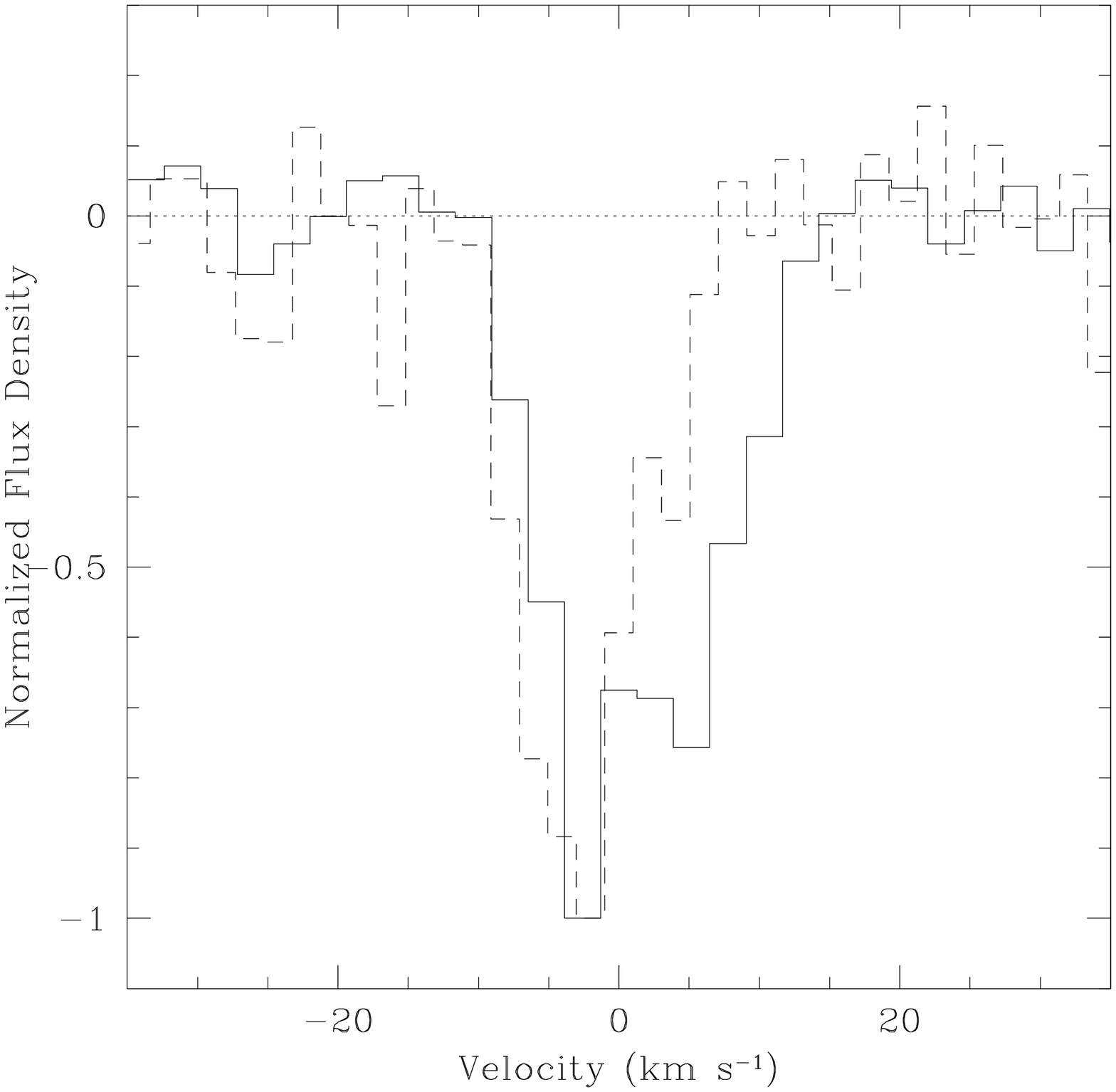,width=5cm,height=6cm}
\hspace{-0.5cm}
\caption{The figure on the right shows a reanalysis of the HI 21cm
absorption spectrum toward 1830-211 made with the 140$'$ telescope [3], 
corrected to optical depth using the 
continuum flux density of 10 Jy. Zero velocity corresponds to
z$_\odot$ = 0.88582. Note that the
channels around zero velocity were corrupted by terrestrial
interference, so only the absorption component near -146 km s$^{-1}$ was
detected. The Gaussian fit parameters for this HI component are:
peak opacity = 0.041$\pm$0.004, and FWHM = 127$\pm$6 km s$^{-1}$
centered at -141$\pm$2 km s$^{-1}$. 
The  WSRT spectrum of Chengalur and de Bruyn [1] shows
HI 21cm absorption at zero velocity with about half the peak optical depth
as that seen at -141 km s$^{-1}$. The figure on the left shows 
redshifted absorption spectra of C$_3$H$_2$ (2$_{12}$-2$_{01}$) 
(solid line) and HC$_3$N (3-2) (dashed line) for the main molecular
absorption line system at z$_\odot$ = 0.88582. The spectra have been
normalized to unit line strength simply to demonstrate the different
velocity structures in the different species.
}
\end{figure}

\vfill
\end{document}